\documentstyle[12pt]{article}
\newcommand{\beq}{\begin{equation}}             %%
\newcommand{\eeq}{\end{equation}}               %%
\newcommand{\bqry}{\begin{eqnarray}}            %%
\newcommand{\eqry}{\end{eqnarray}}              %%
\newcommand{\bqryn}{\begin{eqnarray*}}          %%
\newcommand{\eqryn}{\end{eqnarray*}}            %%
\newcommand{\preprint}[1]{\begin{table}[t]      %%
            \begin{flushright}                  %%
            \begin{large}{#1}\end{large}        %%
            \end{flushright}                    %%
            \end{table}}                        %%
\newcommand{\PD}[2]                             %%
    {\frac{\partial^{#2}}{\partial #1^{#2}}}    %%
\begin{document}
\preprint{LA-UR-98-1695}
\title{Glueball Spectroscopy in Regge Phenomenology\thanks{Presented at the
First International Conference on Parametrized Relativistic Quantum Theory, 
PRQT '98, Houston, Texas, USA, Feb 9-11, 1998}}
\author{\\ L. Burakovsky\thanks{E-mail: BURAKOV@QMC.LANL.GOV} \
\\  \\  Theoretical Division, T-8 \\  Los Alamos National Laboratory \\ Los
Alamos NM 87545, USA}
\date{ }
\maketitle
\begin{abstract}
We show that linear Regge trajectories for mesons and glueballs, and the cubic
mass spectrum associated with them, determine a relation between the masses
of the $\rho $ meson and the scalar glueball, $M(0^{++})=3/\sqrt{2}\;M(\rho ),$
which implies $M(0^{++})=1620\pm 10$ MeV. We also discuss relations between
the masses of the scalar and tensor and $3^{--}$ glueballs, $M(2^{++})=\sqrt{
2}\;M(0^{++}),$ $M(3^{--})=2M(0^{++}),$ which imply $M(2^{++})=2290\pm 15$ 
MeV, $M(3^{++})=3240\pm 20$ MeV.
\end{abstract}
\bigskip
{\it Key words:} Regge phenomenology, glueballs, mesons, mass spectrum

PACS: 12.39.Mk, 12.40.Nn, 12.40.Yx, 12.90.+b, 14.40.Cs 

%\bigskip
\section*{%Introduction
}
The existence of a gluon self-coupling in QCD suggests that, in addition to 
the conventional $q\bar{q}$ states, there may be non-$q\bar{q}$ mesons: bound 
states including gluons (glueballs and $q\bar{q}g$ hybrids). However, the 
theoretical quidance on the properties of unusual states is often 
contradictory, and models that agree in the $q\bar{q}$ sector differ in their
predictions about new states. Moreover, the abundance of $q\bar{q}$ meson 
states in the 1-2 GeV region and glueball-quarkonium mixing makes the
identification of the would-be lightest non-$q\bar{q}$ mesons extremely 
difficult. To date, no glueball state has been firmly established yet.

Although the current situation with the identification of glueball states is 
rather complicated, some progress has been made recently in the $0^{++}$ 
scalar and $2^{++}$ tensor glueball sectors, where both experimental and QCD 
lattice simulation results seem to converge \cite{Land}. Recent lattice 
calculations predict the $0^{++}$ glueball mass to be $1600\pm 100$ MeV 
\cite{Land,Bali,SVW,MP}. Accordingly, there are two experimental candidates 
\cite{pdg}, $f_0(1500)$ and $f_0(1710),$ in this mass range which cannot both
fit into the scalar meson nonet, and this may be considered as strong evidence
for one of these states being a scalar glueball (and the other being dominantly
$s\bar{s}$ scalar quarkonium).

It is known that in both lattice QCD \cite{SVW,Teper,HSZW,LC} and pure 
Yang-Mills \cite{Sam} calculations, the mass of the scalar glueball
is determined by the dimensionless ratio 
\beq
\gamma \equiv \frac{M(0^{++})}{\sqrt{\sigma }},
\eeq
where $\sigma $ is the string tension, and the following values of $\gamma $
have been claimed:
\beq
\gamma =\left[ 
\begin{array}{ll}
3.95, & {\rm ref.}\;[3] \\
4.77\pm 0.05, & {\rm ref.}\;[6], \\
3.88\pm 0.11, & {\rm ref.}\;[8], \\
\approx 3.3, & {\rm ref.}\;[9],
\end{array}
\right.
\eeq
so that the data cluster around $\gamma \simeq 4,$ with uncertainty of $\sim 
15$\%. The use of the value
\beq
\sqrt{\sigma }\approx 0.375\;{\rm GeV},
\eeq
extracted from the light quark Regge phenomenology relation \cite{KS,BB}
\beq
\alpha ^{'}=\frac{1}{8\sigma }\simeq 0.9\;{\rm GeV}^{-2},
\eeq
in Eq. (1) with $\gamma \approx 4$ leads to the following estimate for the 
scalar glueball mass:
\beq
M(0^{++})\simeq 1500\;{\rm MeV}.
\eeq

Here we wish to suggest the following formula for the scalar glueball mass,
\beq
M(0^{++})=3\sqrt{2}\;\!\sqrt{\sigma }\approx 4.24\;\sqrt{\sigma },
\eeq
which, with $\sqrt{\sigma }$ given in (3), predicts
\beq
M(0^{++})\simeq 1600\;{\rm GeV.}
\eeq

A naive way to obtain the formula (6) is to follow the procedure of the 
minimisation of the energy of a bound state of two massless gluons suggested 
by West \cite{West}:
\beq
E=2p+\frac{9}{4}\;\!\sigma r-\frac{\alpha }{r},
\eeq
where $p$ is the gluon momentum, $\alpha $ is the strong coupling constant, 
and 9/4 is a color factor in the long-range confining piece of the two-body 
Coulomb + linear potential. It follows from the uncertainty principle 
$pr\stackrel{>}{\sim }1$ that
\beq
E\geq \frac{2-\alpha }{r}+\frac{9}{4}\;\!\sigma r,
\eeq
and minimising the lower bound of (9) gives
\beq
E=3\sqrt{(2-\alpha )\sigma }\approx 3\sqrt{2}\;\!\sqrt{\sigma }.
\eeq
We note that the procedure of West gives a reasonable result for, e.g., 
ordinary mesons: The analog of (8) in this case is
\beq
E=2\sqrt{p^2+m^2}-2m+\sigma r-\frac{\alpha }{r},
\eeq
where $m$ is the constituent quark mass, and minimising the lower bound of the
corresponding inequality following from (11) leads to the solution (in the 
nonrelativistic approximation $\sqrt{p^2+m^2}\approx m+p^2/(2m))$
\beq
E\simeq 3\left( \frac{\sigma ^2}{4m}\right) ^{1/3}\!\!,\;\;\;r\simeq \left( 
\frac{2}{m\sigma }\right) ^{1/3}\!\!,
\eeq
which, with $\sigma $ given in (3) and $m\simeq 300$ MeV, gives
\beq
E\simeq 765\;{\rm MeV,}
\eeq
in agreement with the $\rho $ meson mass of $\sim 770$ MeV, and $r\simeq 0.7$ 
fm.
  
The way to derive Eq. (6) we suggest here is the use of the hadronic
resonance spectrum. The idea of the spectral description of a strongly 
interacting gas which is a model for hot hadronic matter was suggested by
Belenky and Landau \cite{BL} and consists in considering the unstable 
particles (resonances) on an equal footing with the stable ones in the 
thermodynamic quantities, by means of the resonance spectrum; e.g., the 
expression for pressure in such a resonance gas reads (in the 
Maxwell-Boltzmann approximation)
\beq
p=\sum _ig_i\;p(m_i)=\int _{M_l}^{M_h}dm\;\tau (m)\;p(m),\;\;\;p(m)=\frac{T^2
m^2}{2\pi ^2}K_2\left( \frac{m}{T}\right) ,
\eeq
where $M_l$ and $M_h$ are the masses of the lightest and heaviest species,
respectively, and $g_i$ are particle degeneracies. 

Phenomenological studies \cite{phen} have suggested that the cubic density of 
states, $\tau (m)\sim m^3,$ for each isospin and hypercharge provides a good 
fit to the observed hadron spectrum. Let us demonstrate here that this cubic 
spectrum is intrinsically related to collinear Regge trajectories (for each 
isospin and hypercharge). 

It is very easy to show that the mass spectrum of an individual Regge 
trajectory is cubic. Indeed, consider, e.g., a model linear trajectory with 
negative intercept: 
\beq
\alpha (t)=\alpha ^{'}\;\!t-1.
\eeq
The integer values of $\alpha (t)$ correspond to the states with integer spin, 
$J=\alpha (t_J),$ the masses squared of which are $m^2(J)=t_J.$ Since a 
spin-$J$ state has multiplicity $2J+1,$ the number of states with spin 
$0\leq J\leq {\cal J}$ is
\beq
N({\cal J})=\sum _{J=0}^{\cal J}(2J+1)=({\cal J}+1)^2=\alpha ^{'2}m^4(
{\cal J}),
\eeq
in view of (15), and therefore the density of states per unit mass interval   
(the mass spectrum) is
\beq
\tau (m)=\frac{dN(m)}{dm}=4\alpha ^{'2}m^3.
\eeq
It is also clear that for a finite number of collinear trajectories, the 
resulting mass spectrum is
\beq
\tau (m)=4N\alpha ^{'2}m^3,
\eeq
where $N$ is the number of trajectories, and does not depend on the numerical 
values of trajectory intercepts, as far as its asymptotic form $m\rightarrow
\infty $ is concerned.

It turns out further that the cubic spectrum of the family of collinear 
Regge trajectories enables one to determine the mass of the state this family
starts with. Before we dwell upon this point, let us make the following remark.

It is widely believed that pseudoscalar mesons are the Goldstone bosons of
broken SU(3)$\times $SU(3) chiral symmetry of QCD, and that they should be 
massless in the chirally-symmetric phase. Therefore, it is not clear how well 
would the resonance spectrum be suitable for the description of the 
pseudoscalar mesons. Indeed, as we have tested in \cite{BHNPA}, this nonet is 
{\it not} described by the resonance spectrum. Moreover, pseudoscalar mesons 
are extrimely narrow (zero width) states to fit into a resonance description. 

Thus, the resonance description should start with vector mesons, and the cubic
spectrum of a linear trajectory enables one to determine the mass of the 
$\rho $ meson, as follows:
 
Since the $\rho $ meson has the lowest mass which the resonance description 
starts with, let us locate this state by normalizing the $\rho $ trajectory
to one state in the characteristic mass interval
$(\sqrt{M^2(\rho )-1/(2\alpha ^{'})},\;\!\sqrt{M^2(\rho )+1/(2\alpha ^{'})}).$ 
With the cubic spectrum (17) of a linear trajectory, one has\footnote{Since 
the $\rho $ trajectory starts with a spin-1 isospin-1 state $(\rho ),$ it 
corresponds to the spectrum $\tau (m)=9\times 4\alpha ^{'2}m^3.$ There is 
therefore no difference in normalizing this trajectory to 9 states, or (17) 
to one state, in the vicinity of the $\rho $ mass.}
\beq
1=4\alpha ^{'2}\int _{\sqrt{M^2(\rho )-1/(2\alpha ^{'})}}^{\sqrt{M^2(\rho )+
1/(2\alpha ^{'})}}m^3\;dm=2\alpha ^{'}M^2(\rho ),
\eeq
and therefore 
\beq
M^2(\rho )=\frac{1}{2\alpha ^{'}}.
\eeq
We note that Regge slope is known to have a weak flavor dependence in the
light quark sector \cite{BB}: (in GeV$^{-2})$ $\alpha ^{'}_{n\bar{n}}=0.88,$ 
$\alpha ^{'}_{s\bar{n}}=0.85,$ $\alpha ^{'}_{s\bar{s}}=0.81.$ With $\alpha ^{
'}=0.85$ GeV$^{-2},$ as the average of the above three values, Eq. (20) gives 
\beq
M(\rho )=767\;{\rm MeV,}
\eeq
in excellent agreement with the measured $\rho $ meson mass of $768.5\pm 0.6$
MeV \cite{pdg}. 

It is now tempting to apply similar arguments to glueballs. The slope of 
glueball trajectories can be related to that of meson ones by the product of 
two, color and mechanical, factors:
\beq
\alpha ^{'}_{glue}=\gamma _{c}\gamma _{mech}\;\!\alpha ^{'},
\eeq
where 
\beq
\gamma _c=\frac{4}{9},
\eeq
and
\beq
\gamma _{mech}=\frac{1}{2}.
\eeq
Eq. (23) is easily understood by looking at the confining pieces of the 
two-body potentials in (8),(11), and Eq. (4). The mechanical factor is easily
understood also, by noting that in the string model glueball represents a 
closed string, while ordinary meson an open string, and a spin-energy relation
for a closed string has an extra factor of 1/2 as compared to that for an open
string (e.g., without color, $J=\alpha ^{'}E^2$ for an open string, and $J=1/2
\;\alpha ^{'}E^2$ for a closed string, in the classical case) \cite{Artru}. We
note that the relation $\alpha ^{'}_{glue}=4/9\;\alpha ^{'},$ derived by 
Simonov via the vacuum correlators method \cite{Sim}, capitalizes only the 
color factor, (23), but not the mechanical one, (24). As we see here, both
factors should be taken into account in the correct formula for the glueball 
slope, which is 
\beq
\alpha ^{'}_{glue}=\frac{2}{9}\alpha ^{'}\simeq 0.2\;{\rm GeV}^{-2}.
\eeq 
We now apply a procedure similar to that above for locating the $\rho $ meson 
mass, in order to locate the lightest state which lie on glueball trajectories.
It is firmly established that the lightest glueball state is the scalar 
glueball \cite{West,West2}. We therefore locate the scalar glueball mass as 
that for which its trajectory is normalized to one state in the mass interval 
$(M^2(0^{++})-1/(2\alpha ^{'}_{glue}),\;M^2(0^{++})+1/(2\alpha ^{'}_{glue})).$
Similar to (20), this procedure leads to
\beq
M^2(0^{++})=\frac{1}{2\alpha ^{'}_{glue}},
\eeq
which reduces, through (20),(25), to
\beq
M(0^{++})=\frac{3}{\sqrt{2}}\;\!M(\rho ).
\eeq
We take $M(\rho )=764\pm 5$ MeV, to accommodate the value given in \cite{pdg}
and results on the $\rho ^0$ meson mass, both theoretical \cite{rho-t} and
experimental \cite{rho-x}, which concentrate around 760 MeV. With this 
$M(\rho ),$ the above relation yields
\beq
M(0^{++})=1620\pm 10\;{\rm MeV,}
\eeq
in excellent agreement with QCD lattice results $1600\pm 100$ MeV 
\cite{Land,Bali,SVW,MP}.

We note that it follows from (4),(25),(26) that
\beq
M^2(0^{++})=18\;\!\sigma ,
\eeq
which is equivalent to Eq. (6).

Finally, we note that the mass spectrum can also establish a relation between 
the scalar and tensor glueball mass. Indeed, as discussed in \cite{prqt}, mass
splitting between $S$-wave spin-0 and spin-1 meson states ($\rho $ and $\pi ,$
or $K$ and $K^\ast )$ is well reproduced by the linear and cubic spectra of 
the corresponding multiplets and Regge trajectories, respectively, leading to 
the relations
$$M^2(\rho )-M^2(\pi )=M^2(K^\ast )-M^2(K)=\frac{1}{2\alpha ^{'}},$$
in good agreement with data (first of these relations is consistent with (20),
since $M(\pi )\ll M(\rho )).$ Similar procedure in the glueball sector will 
lead to the relation between the masses of the spin-0 scalar and spin-2 tensor 
glueballs:
\beq
M^2(2^{++})-M^2(0^{++})=\frac{1}{2\alpha ^{'}_{glue}},
\eeq
which implies, through (26),(27),
\beq
M(2^{++})=\sqrt{2}\;\!M(0^{++})=3\;\!M(\rho )=2290\pm 15\;{\rm MeV,}
\eeq
in excellent agreement with QCD lattice simulations \cite{MP,Peard} which give
$2390\pm 120$ MeV for the tensor glueball mass, and corresponding three 
experimental candidates in this mass region \cite{pdg}, $f_J(2220),\;J=2\;{\rm
or}\;4,$ $f_2(2300)$ and $f_2(2340),$ the first of which is seen in $J/\psi 
\rightarrow \gamma +X$ transitions but not in $\gamma \gamma $ production 
\cite{pdg}, while the remaining two are observed in the OZI rule-forbidden 
process $\pi p\rightarrow \phi \phi n$ \cite{pdg}, which favors the gluonium 
interpretation of all three states. It is also interesting to note that, as 
follows from (26),(30), $M^2(2^{++})=1/\alpha ^{'}_{glue},$ and therefore, the
intercept of the tensor glueball trajectory determined by the relation 
$$2=\alpha ^{'}_{glue}M^2(2^{++})+\alpha _{glue}(0)$$ is $\alpha _{glue}(
0)=1.$ This is in agreement with the fact widely accepted in the literature 
that the tensor glueball is the lowest resonance lying on the Pomeron 
trajectory with unit intercept. It is also interesting to note that if one
takes the $3^{--}$ glueball as the Regge recurrence of the tensor glueball
with the above mass value, one will obtain $M^2(3^{--})=2/\alpha ^{'}_{glue},$
and therefore $M(3^{--})=\sqrt{2}\;\!M(2^{++})=2M(0^{++})=3240\pm 20$ MeV, 
consistent with a naive scaling from the two-gluon $2^{++}$ glueball to the 
3-gluon $3^{--}$ case: $M(3g)\simeq 1.5\;\!M(2g)\simeq 3.3$ GeV, with $M(2g)
\simeq 2.2$ GeV. Also, the original constituent gluon model predicts 
$M(1^{--})/M(2^{++})\simeq M(3^{--})/M(2^{++})\simeq 1.5$ \cite{HS}, and QCD
sum rules find the $3g$-glueball mass to be $\simeq 3.1$ GeV \cite{Nar}.

\section*{Concluding remarks}
Again, as in a previous paper \cite{prqt} where we discuss the linear mass
spectrum of an individual hadronic multiplet first established in 
\cite{linear} and then applied to the problem of the correct $q\bar{q}$ 
assignments for problematic meson nonets \cite{BHNPA}, and the cubic spectrum 
of a strongly interacting gas as a model for hot hadronic matter, we have 
again demonstrated that a mass spectrum represents a powerful tools for hadron
spectroscopy. We have shown that linear trajectories for mesons and glueballs,
and the cubic mass spectrum associated with them, determine a relation between
the masses of the lightest states lying on these trajectories, the $\rho $ 
meson and the scalar glueball, respectively, which implies that the mass of 
the latter is in the vicinity of 1600 MeV. We have also established a relation
between the scalar glueball mass and string tension (Eq. (29)) which is the 
subject of lattice QCD and pure Yang-Mills calculations, and discussed 
relations between the scalar and tensor and $3^{--}$ glueball masses.

\section*{Acknowledgements}
Correspondence with L.P. Horwitz during the preparation of this work is very
much appreciated.

\end{document}